\documentclass[conference,a4paper]{IEEEtran}
\IEEEoverridecommandlockouts
\usepackage{cite}
\usepackage{amsmath,amssymb,amsfonts}
\usepackage{algorithmic}
\usepackage{graphicx}
\usepackage{textcomp}
\usepackage{xcolor}
\def\BibTeX{{\rm B\kern-.05em{\sc i\kern-.025em b}\kern-.08em
    T\kern-.1667em\lower.7ex\hbox{E}\kern-.125emX}}

\usepackage{mathtools}
\usepackage{bm}
\usepackage{multirow}
\usepackage{bbding}
\usepackage[hidelinks]{hyperref}
\usepackage{cleveref}
\usepackage{dsfont}
\usepackage{algorithm}

\newcommand{\R}{\mathbb{R}}

\begin{document}

\title{The Neural Echo: A Signal Processing Perspective\\ 
for Understanding Neural Networks 
\thanks{C.W.~and D.G.~contributed equally to this work.}
\thanks{C.W.~ is supported by the Konrad Zuse School of Excellence in Learning 
 and Intelligent Systems (\href{https://eliza.school/}{ELIZA}) through the DAAD 
 programme ``Konrad Zuse Schools of Excellence in Artificial Intelligence'', 
 sponsored by the Federal Ministry of Education and Research.}
 \thanks{We gratefully acknowledge the stimulating research environment of the 
 GRK 2853/1 ``Neuroexplicit Models of Language, Vision, and Action''
 funded by the Deutsche Forschungsgemeinschaft (DFG, German Research 
 Foundation) under project number 471607914.}
 \thanks{\copyright~2026 IEEE. Personal use of this material is permitted. Permission from IEEE must be obtained for all other uses, in any current or future
media, including reprinting/republishing this material for advertising or
promotional purposes, creating new collective works, for resale or
redistribution to servers or lists, or reuse of any copyrighted
component of this work in other works.}
}

\author{
\IEEEauthorblockN{Chongbiao Wang\IEEEauthorrefmark{1}\IEEEauthorrefmark{2},
    Daniel Gaa\IEEEauthorrefmark{1},
    Joachim Weickert\IEEEauthorrefmark{1},
    Karl Schrader\IEEEauthorrefmark{1}}
\IEEEauthorblockA{
    \IEEEauthorrefmark{1}Mathematical Image Analysis Group \\
    Faculty of Mathematics and Computer Science, Saarland University \\ 
    Campus E1.7, 66041 Saarbr\"ucken, Germany\\
    \{cwang, gaa, weickert, schrader\}@mia.uni-saarland.de\\
    \IEEEauthorrefmark{2}Zuse School ELIZA \\ 
    Hochschulstr. 10, 64289 Darmstadt, Germany}
}

\maketitle

\begin{abstract}
We introduce the neural echo as a tool for understanding the behavior of 
neural networks. It generalizes the model-based concepts of impulse 
responses, diffusion echoes, and filter echoes to learning-based 
methods. It provides local, space-adaptive impulse responses and filter
kernels for a neural network, its so-called echoes. These echoes depend on
the input image and can be visualized to understand the learned dynamics of 
the network via an affine mapping. Neural echoes build a bridge from 
classical signal processing to modern explainable AI. They are very 
general and can be applied to both image-to-image and classification 
networks, with convolutional or fully connected structure, of feedforward 
or recurrent type, including modern transformer networks. Network
differentiability is not required. In the differentiable case, neural 
echoes comprise concepts based on the network Jacobian, such as saliency 
maps and the analysis of adversarial perturbations, as special instances. 
As a simple blueprint to explain our framework, we derive neural echoes 
for the denoising convolutional neural network (DnCNN). Our experiments 
suggest that this network weights pixels based on their spatial and 
gray value distances. This not only clarifies its behavior, but also 
shows that it can reproduce key concepts of classical model-based 
denoisers such as bilateral filtering. 
\end{abstract}

\begin{IEEEkeywords}
Signal Processing, Neural Networks, Impulse Response
\end{IEEEkeywords}
%
%

\section{Introduction}

The {\em impulse response} of a linear shift-invariant (LSI) filter is 
a key concept in classical signal processing \cite{OSB99}. It is obtained 
by applying the filter to an impulse signal and transparently expresses 
the action of the LSI filter at every position. Unfortunately, their lack 
of adaptivity to the signal makes LSI filters unsuitable for more complex 
filtering tasks.

Shift-variant and potentially nonlinear filters are more versatile, but
their space-adaptive behavior means that no classical impulse response 
can be formulated for them. In 2001, Dam and Nielsen~\cite{DN01} introduced 
the {\em diffusion echo}. Their source diffusion echo can be seen as a 
space-variant impulse response for nonlinear diffusion filters. Recently, 
this concept has been extended to general model-based image processing 
and computer vision filters by Gaa et al.~\cite{GWFC26}, who have 
established the notion of the {\em filter echo}. They show that 
the action of many common filters can be expressed in terms of a 
matrix-vector multiplication of the input image with a state transition 
matrix. Its columns are the space-adaptive source echoes of the filter. 
Moreover, the rows of the matrix coincide with the definition of the 
drain echo from~\cite{DN01}. Since these model-based filters rely on 
compact mathematical formulations, they offer a certain degree of 
interpretability by design. However, for more sophisticated filters 
involving e.g.~partial differential equations (PDEs) or variational 
methods, the specific filter effects may not always be obvious from 
their mathematical formulations alone. The diffusion and filter echoes 
provide an intuitive visual interpretation of the filter action in 
terms of a clear signal processing-based concept: a simple linearized 
characterization via a matrix-vector product. This allows for the 
comparison of different filters in a maximally transparent way and 
underlines the usefulness of these echoes. 

These days, digital image analysis is dominated by learning-based 
approaches that rely on deep neural networks~\cite{GBC16,LBH15}. 
However, neural networks are still largely treated as black-box 
filters. Attempts to understand their inner workings are a hot 
research topic that has created the field of {\em explainable AI}.

So far, the filter echo framework has been applied exclusively to 
model-based approaches~\cite{GWFC26}. Thus, the natural question arises 
whether a neural network can be reformulated to comply with this framework. 
This would enable an interpretation of neural networks from a signal 
(or image) processing perspective and thus build a bridge between two 
important fields: classical signal processing and explainable AI.


\smallskip
\noindent
{\bf Our Contributions.}
The present work closes this gap by establishing a generalization of 
the filter echo framework from model-based approaches to neural networks. 
To this end, we introduce the \emph{neural echo}. We show how one can 
rewrite the inference phase of a neural network such that it fits 
(a generalization of) the framework in~\cite{GWFC26} and yields 
filter echoes. 

As a blueprint, we present results for image denoising with a 
well-established neural approach: the DnCNN~\cite{ZZC+17} of
Zhang et al.~\cite{ZZC+17}.
We do not consider more complex architectures or filter tasks, 
since all concepts and foundational insights can already be 
obtained from this simple model. We compare its neural echoes 
to bilateral filter echoes~\cite{AW95,SB97,TM98}. This enables a 
deeper understanding of the neural network and emphasizes its 
connections to classical image processing.

Having shown how neural echoes are obtained for this denoising
network, we discuss the framework in its full generality. We sketch how
other typical network components can be included. Our framework 
covers both image-to-image and classification networks,
convolutional or fully connected, of feedforward or recurrent type. 
Transformer networks are also covered, and no differentiability
assumptions are required.

To establish a bridge to existing explainable AI methods, we show 
that for differentiable networks, approaches based on the network 
Jacobian are special instances of our framework. This
includes e.g.~saliency maps~\cite{SVZ14} and the analysis of
adversarial perturbations~\cite{PMJ+16,SZSB14}. 
It offers an alternative, signal processing-inspired viewpoint on 
these popular concepts.
 

\smallskip
\noindent
{\bf Related Work.}
Related ideas exist for network visualization, which have been studied 
mainly for image classification. The so-called saliency maps of 
Simonyan et al.~\cite{SVZ14} are built using partial
derivatives of output neurons w.r.t.~input neurons. Thus, they visualize 
the influence of changes in pixels in the input image on the classification 
decision; see also~\cite{SDBR15,ZF14}. 
Network derivatives also play a natural role in adversarial 
perturbations, where the effect of a change in an input pixel on the 
classification output is studied~\cite{PMJ+16,SZSB14}.
In \Cref{sec:general}, we will see that tools based on network derivatives 
can be regarded as special instances of our neural echo framework.

Since we deal with a neural denoising method, we refer to~\cite{EKV23}
for a review on such approaches. Classical model-based denoising methods 
have been around for a long time; see Lebrun et al.~\cite{LCBM12} for 
an overview and a principled analysis. 


\smallskip
\noindent
{\bf Paper Organization.}
We start with brief reviews of the filter echo and the DnCNN denoiser 
in \Cref{sec:filter-echo} and \Cref{sec:dncnn}.
In \Cref{sec:neural-echo}, we show how to fit such a neural network 
into the filter echo setting, which allows us to introduce the neural 
echo. The generality of our approach and its relation to concepts based
on the network Jacobian are discussed in \Cref{sec:general}.
We present a denoising experiment in \Cref{sec:experiments} and conclude 
the paper in \Cref{sec:conclusion}.


\section{A Review of the Filter Echo Framework}
\label{sec:filter-echo}

The filter echo framework has been introduced in~\cite{GWFC26} as an 
extension of the diffusion echo~\cite{DN01}. For a discrete input image 
stacked into a vector $\bm{f}\in \R^N$, the framework describes the 
filter output $\bm{u} \in \R^M$ by a matrix-vector product with a 
suitable state transition matrix $\bm{S} \in \R^{M \times N}$:
\begin{equation}
    \label{eq:basic}
    \bm{u} = \bm{S} \bm{f} \,.
\end{equation}

This is a fairly common formulation in image filtering~\cite{Mi13a} and 
subsumes many well-established filters. It is important to note that the 
matrix $\bm{S}=\bm{S}(\bm{f})$ may depend on the input image~$\bm{f}$. 
Since its entries can result from nonlinear adaptations of the filter to 
$\bm{f}$, \Cref{eq:basic} can also cover very complex filters. 
By writing a highly sophisticated nonlinear filter like it were a specific 
space-variant linear filter, it offers an intuitive and unified 
interpretation of various filters.

Based on the definition in~\cite{DN01}, one distinguishes between a 
source echo and a drain echo. The source echo can be interpreted as a 
space-varying impulse response of the filter. As such, the source echo at a 
location $i$ describes the propagation of a gray value $f_i$ through the 
filtering process. On the other hand, the drain echo is the local filter 
kernel that tells us from where the gray value information $u_i$ at the 
corresponding location in the filtered image originated. Thus, the source 
and drain echoes describe the filter from the perspective of the sender 
and the receiver, respectively.

In~\cite{GWFC26}, it is shown that the source echo~$\bm{s}_i$ at 
location $i$ corresponds to the $i$-th column of the matrix~$\bm{S}$, 
while the drain echo~$\bm{d}_i$ is given by its $i$-th row. Therefore,
we have
\begin{align}
 \label{eq:source}
 \bm{s}_i &= \bm{S}\,\bm{e}_i\,,\\
 \label{eq:drain}
 \bm{d}_i &= \bm{S}^\top\!\bm{e}_i\,,
\end{align}
where $\bm{e}_i$ is a unit impulse image, which is $1$ in pixel $i$ 
and $0$ elsewhere. In general, $\bm{S}$ is nonsymmetric and the source 
and drain echoes differ. The sets of drain or source echoes in all pixels 
contain the full information on the filtering process and, as such, 
allow to reconstruct the filtered image. 

It is important to note that the matrix~$\bm{S} \in \R^{M \times N}$ 
can be very large. Therefore, it is usually too expensive to explicitly 
compute and/or store the entire matrix in a naive way. In the examples 
in~\cite{GWFC26}, this matrix is either sparse and can be directly 
deduced from the model weights or obtained by (repeatedly) 
solving linear systems of equations with sparse system matrices. Therefore, 
in these cases, the matrix $\bm{S}$ can be applied implicitly without 
computing the full matrix. We also make use of this idea for our neural 
echoes.


\section{A Review of the DnCNN}
\label{sec:dncnn}

The DnCNN by Zhang et al.~\cite{ZZC+17} is a denoising approach based on 
residual learning~\cite{HZRS16}. This means that it learns to 
predict the noise profile, which is then subtracted from the noisy 
input image to obtain the clean result.

Let $\{\widetilde{\bm{u}}_{\bm{k}}, \bm{f_k}\}$ be a discrete training image 
pair with a clean image $\widetilde{\bm{u}}_{\bm{k}} \in \R^N$ and a noisy 
image $\bm{f_k} \in \R^N$. 
Then the image $\bm{u_k} \in \R^N$ denoised by the DnCNN is given by
\begin{equation}
  \label{eq:dncnn-prediction}
  \bm{u_k} = \bm{f_k} - \mathcal{R}_{\bm{\theta}^\ast} (\bm{f_k})\,,
\end{equation}
where $\mathcal{R}_{\bm{\theta}^\ast}$ is the network operator that 
produces the predicted residual from the noisy input $\bm{f_k}$.
The optimized network parameters $\bm{\theta}^\ast$ are obtained by 
minimizing the mean squared error between clean and denoised images 
in the training set.

\begin{figure}[!tb]
\centering
\includegraphics[width=0.96\linewidth]{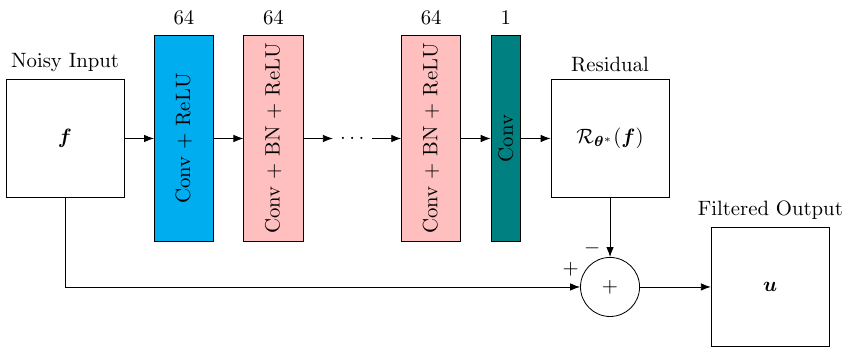}

\vspace{-1mm}
\caption{\label{fig:architecture} The DnCNN architecture.
Adapted from~\cite{ZZC+17}.}
\end{figure}

\Cref{fig:architecture} shows the complete architecture of the DnCNN.
It consists of $D\geq2$ convolutional layers, where all but the final 
convolution are followed by a ReLU activation function~\cite{Ho41,NH10}.
Furthermore, the inner $D-2$ layers also contain a batch normalization (BN)
layer~\cite{IS15}.

The convolution in the first layer produces $64$ feature maps of the same 
size as the input image. The convolution in each inner layer keeps the 
number and size of the feature maps constant, while the last layer 
creates the output residual image from its $64$ input feature maps.


\section{The Neural Echo}
\label{sec:neural-echo}

To explain the basics behind the neural echo, we consider a simple
proof-of-concept application: denoising of additive Gaussian noise
with the DnCNN. For computing the neural echoes, we only need the 
inference phase of the network. Thus, we assume that the network
parameters have already been optimized on some training set and 
remain fixed. From now on, we refer to the network operator simply 
as $\mathcal{R}$.

To apply the filter echo framework, we need to rewrite the process 
\labelcref{eq:dncnn-prediction} of producing a denoised image as
\begin{equation}
    \label{eq:filter-formulation-nn}
    \bm{u} = \bm{S} \,\bm{f} \, ,
\end{equation}
with a matrix $\bm{S} \in \R^{N \times N}$ that depends on the input image
$\bm{f}$. 

We now discuss how individual components of the DnCNN can be written as
matrix-vector multiplications. Additive terms, such as biases in 
convolutional layers or subtraction of the mean in the BN layers, require
extra care. Therefore, we begin with a simplified version of the 
DnCNN without bias terms and BN layers and show how to write it in 
the form of \labelcref{eq:filter-formulation-nn}. 
Afterwards, we extend the filter echo framework so that it also subsumes the 
full DnCNN.


\subsection{Simplified DnCNN without Biases and Batch Normalization}


\noindent
{\bf Convolutional Layer.} 
A convolutional layer (without bias) can be written as a matrix-vector 
multiplication by appropriately arranging the kernel weights across the rows 
of the filter matrix. The resulting sparse convolution matrix incorporates 
the boundary conditions. It is applied to the vectorized image to obtain the 
convolved result. In the case of the DnCNN, which uses a $3 \times 3$ filter 
kernel, the resulting matrix has at most nine nonvanishing entries per row.
The 64 feature maps of the inner layers of the DnCNN can be stacked into a 
vector of size $64N$ such that the inner convolution matrices have 
dimension $64N \times 64N$, while the convolution matrices of the first and
last layers are of dimension $64N \times N$ and $N \times 64N$, respectively. 
We denote the convolution matrices by $\bm{K}$.


\smallskip
\noindent
{\bf Rectified Linear Unit.}
The Rectified Linear Unit (ReLU)~\cite{Ho41,NH10} is an activation 
function given by $\mathrm{ReLU}(x) \!=\! \max\,(0, x)$.
Applying it to every component $x_i$ of a vector $\bm{x}$ 
comes down to multiplying $\bm{x}$ with a diagonal matrix 
${\bm{R}(\bm{x})}=(r_{i,j})$. It satisfies $r_{i,i} = 1$ if $x_i > 0$, 
and $0$ otherwise.


\smallskip
\noindent
{\bf Filter Echo Formulation of the Simplified Network.}
To write the entire simplified network as matrix--vector multiplication,
we concatenate the matrices of its layers. The predicted residual 
output of a $D$-layer network for a given input image $\bm{f} \in \R^N$ 
can be written as
\begin{equation}
    \mathcal{R} (\bm{f}) \,=\, 
    \underbrace{\bm{K}^{(D)} \bm{R}^{(D-1)} \bm{K}^{(D-1)} \cdots 
    \bm{R}^{(1)} \bm{K}^{(1)}}_{\bm{P}(\bm{f})}\, \bm{f} \,,
\end{equation}
where the upper indices of the matrices refer to the layers and where 
each ReLU matrix depends on its input. The matrix~$\bm{P}$ is the product 
of sparse convolution matrices and diagonal ReLU matrices. It depends only
on the input image $\bm{f}$, since the trained convolution weights are 
fixed during inference. 

The denoised image $\bm{u}$ is obtained by subtracting the predicted 
residual from the noisy image:
\begin{equation}
  \bm{u} \,=\, \bm{f} - \mathcal{R} (\bm{f}) 
         \,=\, \underbrace{(\bm{I} - \bm{P}(\bm{f}))}_{\bm{S}(\bm{f})} 
               \, \bm{f}.
\end{equation}

This formulation yields the state transition matrix $\bm{S} \in 
\R^{N \times N}$, which maps the noisy input to the filtered output. 
Its columns and rows constitute the echoes that correspond to the action of 
the network when applied to the specific input image. It is interesting to 
note that in contrast to echoes of other filters, for example diffusion or 
inpainting~\cite{GWFC26}, neural echoes are not restricted to being 
probability distributions and can have arbitrary, potentially 
negative components.

Since the matrices $\bm{S}$, $\bm{K}^{(k)}$, and $\bm{R}^{(k)}$ can become
prohibitively large, it is not recommended to explicitly compute or store them. 
Instead, echoes can be extracted using (\ref{eq:source}) or (\ref{eq:drain}) 
and applying the individual matrices through the implementations of the 
convolution and ReLU layers of a deep learning framework. Since the ReLU 
activation matrices $\bm{R}^{(k)}$ depend on the input of the layers,
the implementation of the ReLU layer needs to be slightly adapted when computing
the echoes. A pseudocode of the algorithm for computing neural source echoes for
the simplified network is given in \Cref{alg:echo-computation}. For the 
corresponding drain echoes, we can apply the layers in reverse order and use 
transposed convolutions.


\begin{algorithm}[tbhp]
\caption{Neural Source Echo Computation for the Simplified Network. 
\label{alg:echo-computation}}
\small
\begin{description}
\hrule\vspace{2mm}
\item[\emph{Input:}]\hfill\\[0.5mm]
     Input image $\bm f \in \R^{N}$,\\
     trained simplified DnCNN architecture with $D$ layers,\\
     index set $J \subset \{1, \dots, N\}$ of desired source echo locations.
\vspace{-2.5mm}
\item[\emph{Initialization:}]\hfill\\[0.5mm]
     Initialize residual with initial image: $\bm{r}^{(0)} = \bm{f}$.
     Initialize residual echoes with unit impulses: 
     $\,\bm{q}_{i}^{(0)} = \bm{e}_i$\, for $i \in J$.
\vspace{1.5mm}
\item[\emph{Compute:}]\hfill\\[0.5mm]
     \textbf{for} $k = 1, \dots, D-1$ 
     \begin{enumerate}
     \item[1.] Apply convolution matrix of current layer to residual and 
               residual echoes to obtain intermediate results:\\[1mm]
               $\widetilde{\bm{r}}^{(k)} = \bm{K}^{(k)} \bm{r}^{(k-1)}$,\, 
               $\widetilde{\bm{q}}_{i}^{(k)} 
               = \bm{K}^{(k)} \bm{q}_{i}^{(k-1)}$\, for $i \in J$.
     \vspace{-2.5mm}
     \item[2.] Apply ReLU to intermediate residual results:\\[1mm]
               $\bm{r}^{(k)} = \bm{R}^{(k)}(\widetilde{\bm{r}}^{(k)}) \, 
               \widetilde{\bm{r}}^{(k)}$.
     \vspace{1mm}
     \item[3.] Apply \emph{same} ReLU to intermediate residual echoes:\\[1mm]
               $\,\bm{q}_{i}^{(k)} = \bm{R}^{(k)} (\widetilde{\bm{r}}^{(k)}) 
               \, \widetilde{\bm{q}}_{i}^{(k)}$\, for $i \in J$.
     \vspace{0.5mm}
     \end{enumerate}
     \textbf{end for}\\[1mm]
     Apply convolution of output layer:\\
     $\bm{r} = \bm{K}^{(D)} \bm{r}^{(D-1)}\,$ and
     $\,\bm{q}_{i} = \bm{K}^{(D)} \bm{q}_{i}^{(D-1)}$\, for $i \in J$.\\[1mm]
     Subtract residual from input: $\bm{u} = \bm{f} - \bm{r}$.\\ 
     Equivalently for echoes: 
     $\,\bm{s}_i = \bm{e}_i - \bm{q}_i$\, for $i \in J$.
\vspace{1.5mm}
\item[\emph{Output:}]\hfill\\[0.5mm]
     Denoised image $\bm{u} \in \R^N$,\\
     source echoes $\bm{s}_i \in \R^N$ for $i \in J$.\\
     \vspace*{-1mm}\hrule
     \end{description}
\end{algorithm}


\subsection{Full DnCNN}

The full DnCNN~\cite{ZZC+17} also contains bias terms and batch normalization 
layers. These terms complicate our considerations, since additive terms are 
not covered by the filter echo framework. We explain how to deal with this in 
the following.


\smallskip
\noindent
{\bf Bias Term.}
For an input $\bm{f} \in \R^N$ to a convolutional layer with weight matrix 
$\bm{K} \in \R^{N \times N}$ and bias vector $\bm{b} \in \R^N$, 
the output $\bm{u} \in \R^N$ is
\begin{equation}
    \bm{u} = \bm{K}\bm{f} + \bm{b} \,.
\end{equation}

To write this as a single matrix-vector multiplication, we use the 
following matrix extension that is popular e.g.~for homogeneous coordinates 
in geometric computer vision~\cite{Fu24}:
\begin{equation}
    \underbrace{\begin{pmatrix}
        \bm{u} \\
        1
    \end{pmatrix}}_{\widehat{\bm{u}}}
    = 
    \underbrace{\begin{pmatrix}
        \bm{K} & \bm{b} \\
        \bm{0}^\top & 1
    \end{pmatrix}}_{\widehat{\bm{K}}}
    \underbrace{\begin{pmatrix}
        \bm{f} \\
        1
    \end{pmatrix}}_{\widehat{\bm{f}}}.
\end{equation}


\smallskip
\noindent
{\bf Batch Normalization.}
In the inference phase, batch normalization (BN)~\cite{IS15} affinely 
transforms the elements of the input vector. The weights of this 
transformation are fixed based on the statistics of the training 
set. Although the multiplicative part of the normalization is 
straightforwardly expressed with a diagonal matrix, the additive part poses 
the same problem as the bias term in the convolutional layer. Thus,
as a remedy, we also introduce an extended matrix formulation of a batch 
normalization layer and denote it by $\widehat{\bm{B}}$.


\smallskip
\noindent
{\bf Filter Echo Formulation of the Full Network.}
We can again express the full network by concatenating the extended 
matrices of all layers. Eventually, the denoised image $\bm{u} \in \R^{N}$ 
is simply extracted by considering the first $N$ entries of the result 
$\widehat{\bm{u}} \in \R^{N+1}$.

To fit the other layers, the extended ReLU matrix reads
\begin{equation}
    {\widehat{\bm{R}}} = 
    \begin{pmatrix}
        \bm{R} & \bm{0} \\
        \bm{0}^\top & 1
    \end{pmatrix}\,.
\end{equation}
This allows to express the full DnCNN from Fig.~\ref{fig:architecture} as
\begin{equation}
    \widehat{\mathcal{R}}(\bm{f}) 
    \,=\, \underbrace{\widehat{\bm{K}}^{(D)} \widehat{\bm{R}}^{(D-1)} 
    \widehat{\bm{B}}^{(D-1)} \widehat{\bm{K}}^{(D-1)} \cdots 
    \widehat{\bm{R}}^{(1)} \widehat{\bm{K}}^{(1)}}_
    {\widehat{\bm{P}}\left(\bm{f}\right)} \,\widehat{\bm{f}} \, .
\end{equation}
By induction it follows that $\widehat{\bm{P}}({\bm f})$ 
has the structure 
\begin{equation}
\widehat{\bm{P}} \,=\,  
    \begin{pmatrix}
        \bm{P} & \bm{d} \\ \bm{0}^\top & 1
    \end{pmatrix}.
\end{equation}
This gives
\begin{equation}
 \label{eq:extended-full}
 \widehat{\mathcal{R}}(\bm{f}) \,=\, \widehat{\bm{P}} \widehat{\bm{f}} \,=\,
    \begin{pmatrix}
        \bm{P} & \bm{d} \\ \bm{0}^\top & 1
    \end{pmatrix}
    \begin{pmatrix}
        \bm{f} \\ 1
    \end{pmatrix}
    \,=\,
    \begin{pmatrix}
        \bm{P} \bm{f} \\ 1
    \end{pmatrix}
    +
    \begin{pmatrix}
        \bm{d} \\ 1
    \end{pmatrix}.
\end{equation}
Thus, the residual $\bm{r}$ satisfies 
\begin{equation}
    \label{eq:residual-ext}
    \bm{r} = \bm{P} \bm{f} + \bm{d}.
\end{equation}

The matrix $\bm{P}$ contains the multiplicative components of the layers
(i.e.~the convolution without bias, the multiplicative part of BN and the
ReLU activation). The vector $\bm{d}$ comprises the additive terms, i.e. 
the sum of all individual biases, each multiplied by all 
multiplicative components that follow it.

The denoised image is given by
\begin{equation}
 \widehat{\bm{u}} 
    \,=\, \widehat{\bm{f}} - \widehat{\mathcal{R}}(\bm{f}) 
    \,=\, \underbrace{\left(\widehat{\bm{I}} - \widehat{\bm{P}}
          \left(\bm{f}\right)\right)}_{\widehat{\bm{S}}\left(\bm{f}\right)} 
          \, \widehat{\bm{f}}\,,
\end{equation}
which comes down to 
\begin{equation}
    \bm{u} = (\bm{I} - \bm{P}) \bm{f} - \bm{d}.
\end{equation}

To compute the echoes of the full DnCNN, we can first compute the echoes of 
the multiplicative parts analogously to \Cref{alg:echo-computation}, but 
setting all the additive parts of the layers to zero. Finally, we compute 
the bias echo $-\bm{d}$ by starting with the bias of the first convolutional 
layer and applying all subsequent layers including their additive terms.


\section{Neural Echoes for General Networks} 
\label{sec:general}

\noindent
{\bf General Setting.}
The reason why we have introduced neural echoes for the DnCNN denoiser 
-- a relatively simple image-to-image network -- was to explain the 
foundational ideas with a transparent example. Having mastered these
technicalities, it is straightforward to extend neural echoes 
to a general setting where the size $N$ of the input data 
$\bm{f} \in \R^N$ differs from the dimension $M$ of the network output 
$\bm{u} \in \R^M$. In homogeneous coordinates, we obtain 
\begin{equation}
 \label{eq:general}
 \bm{\widehat{u}} \,=\, \bm{\widehat{S}}(\bm{\widehat{f}})\,\bm{\widehat{f}}
\end{equation} 
with an extended state transition matrix 
$\bm{\widehat{S}} \in \R^{(M+1)\times(N+1)}$.
This has interesting and far-reaching consequences, e.g.:
\begin{itemize}
\item It applies not only to image-to-image networks with 
      different input and output dimensions (e.g.~for optical flow 
      computations), but also to classification networks. 
\item If $\bm{\widehat{S}}$ is the product of individual matrices, they 
      do not have to be quadratic. Thus, they may e.g.~represent up- or 
      downscaling operations such as max-pooling~\cite{WAH92,BBLP10}.
\item The setting is not restricted to convolutional networks that share
      their weights within one layer. We can consider general fully 
      connected networks without weight sharing. 
\item Even modern architectures such as transformers~\cite{VSP+17} 
      are covered by our framework.
      Crucially, self-attention acts as an input-dependent weighting
      matrix, where the data dynamically dictate the linear combination
      of values.
\item Our framework also incorporates recurrent networks, since they can
      be unrolled to feedforward ones~\cite{MLE21}.
\end{itemize}
Thus, we can cover very general neural network architectures. 


\smallskip
\noindent{\bf Coverage of Saliency Maps and Adversarial Perturbations.}
Several existing concepts visualize and analyze classification
networks via first-order partial derivatives (i.e.~the network 
Jacobian). This includes ideas from explainable AI such as saliency 
maps~\cite{SDBR15,SVZ14,ZF14} and adversarial 
perturbations~\cite{PMJ+16,SZSB14}. 
Let us now interpret such approaches from our neural echo perspective. 
Expressing (\ref{eq:general}) in regular rather than homogeneous 
coordinates, we obtain an affine transformation 
\begin{equation}
 \label{eq:general2}
 \bm{u} \,=\, \bm{S}(\bm{f})\,\bm{f} + \bm{b}
\end{equation} 
that characterizes the network output $\bm{u} \in \R^M$ for some input 
data $\bm{f} \in \R^N$ in terms of the state transition matrix
$\bm{S}(\bm{f}) \in \R^{M \times N}$ and a suitable offset vector 
$\bm{b} \in \R^M$. For a differentiable neural network, a Taylor 
linearization gives the same formula, with $\bm{S}(\bm{f})$ being the 
Jacobian of the network; see also~\cite{SF19,WMC+16}. This naturally 
comprises saliency maps. Since they compute partial derivatives of 
output neurons w.r.t.~input neurons, they rely on neural drain echoes.
Furthermore, the identification of $\bm{S}(\bm{f})$ with
the Jacobian implies that the columns of the Jacobian are the source 
echoes of the network. This takes the perspective of adversarial attacks, 
where the effect of a change in an input pixel on the classification 
output is studied~\cite{PMJ+16,SZSB14}. 

Our discussion shows that for {\em differentiable} networks, visualization
and analysis tools which involve the network Jacobian can be treated
within the neural echo framework. This supplements them with a novel 
interpretation from a signal processing viewpoint. 
We would like to stress, however, that our framework is more general: 
Since it does not require to propagate derivatives through the network, 
it can also characterize {\em nondifferentiable} behavior, both from 
a sender's (neural source echo) and a receiver's (neural drain echo) 
perspective.
 

\section{Denoising Experiment}
\label{sec:experiments}

\begin{figure}[!t]
\centering
\tabcolsep2pt
\begin{small}
\begin{tabular}{ccc}
input image & DnCNN result & DnCNN echo \\[0.5pt]
\includegraphics[width=0.3\linewidth]
  {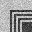} &
\includegraphics[width=0.3\linewidth]
  {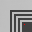} &
\includegraphics[width=0.3\linewidth]
  {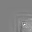} \\[1.5pt]
&
\includegraphics[width=0.3\linewidth]
  {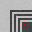} &
\includegraphics[width=0.3\linewidth]
  {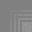} \\[0.5pt]
  & bilateral result & bilateral echo
\end{tabular}
\end{small}

\caption{\label{fig:ex1} Comparison of source echoes of the DnCNN
and the bilateral filter with optimized parameters ($\sigma_s = 12.5$,
$\sigma_t = 43$) on a noisy synthetic test image.
The echoes are scaled such that the maximum absolute 
value is mapped to 0 or 255 and weights of zero are displayed in gray 
$(127.5)$. The red dot in the filtered images marks the echo location.}
\end{figure}

As a proof-of-concept application, we now present a denoising experiment. 
Its purpose is not a comprehensive empirical evaluation, but to illustrate 
the conceptual utility of neural echoes in a simple and transparent setting. 
We have trained the DnCNN for 
additive Gaussian noise with zero mean and standard deviation $\sigma=15$. 
Like in~\cite{ZZC+17}, we employ 17 network layers.
Since we aim to visualize the behavior of the network on a foundational 
level, we use a simplistic binary test image in \Cref{fig:ex1}, 
inspired by the experiments in~\cite{GWFC26}.

Consequently, we also train the network on binary data. We choose the same 
$400$ training images as~\cite{ZZC+17}, but binarize the resulting patches 
with Otsu thresholding~\cite{Ot79}. Like in~\cite{ZZC+17}, we use a 
batch size of $128$ and crop $1600 \times 128$ patches of size 
$40 \times 40$ that we augment with random rotations to avoid 
directional bias. We employ the Adam optimizer~\cite{KB15} with a 
learning rate of $0.001$ to train for $60$ epochs, during which we 
exponentially decrease the learning rate with a factor of $0.95$. 

In \Cref{fig:ex1}, we add Gaussian noise with $\sigma=15$ to our 
test image of size $32 \times 32$ and range $[0,255]$. We denoise 
it with the DnCNN and compare to a bilateral filter~\cite{AW95,SB97,TM98}. 
The spatial and tonal (i.e.~grayscale) standard deviations $\sigma_s$ 
and $\sigma_t$ of its Gaussian weight functions are optimized w.r.t.~the 
mean squared error (MSE). \Cref{fig:ex1} shows an exemplary source 
echo that visualizes the filter behavior of both methods.

We see that the bilateral filter can ``jump'' to distant segments, 
if they are tonally similar (i.e. have similar gray values). Moreover, 
we observe that its Gaussian decay of the spatial weight assigns higher 
relative weights to closer pixels. 

The echo of the neural network illustrates that it has also learned the 
concept of spatial and tonal similarity. This demystifies its behavior
and confirms the usefulness of these handcrafted key components of the 
bilateral filter. Interestingly, the DnCNN MSE of $2.08$ is substantially 
lower than the bilateral filter MSE of $15.46$. Moreover,
the DnCNN echo appears to show a stronger tonal adaptation to the 
data. This indicates how one could improve the explicit bilateral 
filter model by modifying its tonal weight function.

We observe that already such a simple experiment with the neural echo 
and a model-based filter echo provides valuable insights from which both 
worlds may benefit.  


\section{Conclusions and Outlook}
\label{sec:conclusion}

We have introduced the \emph{neural echo} as a general concept 
for analyzing and understanding the filter effect of neural networks. 
To this end, we have written 
the action of each network layer as a matrix--vector 
product with matrix entries that may even subsume the outcome of 
nonlinear mappings, e.g.~by activation functions. To also incorporate 
affine transformations caused by bias terms or batch normalizations, 
we have supplemented our setting with extended matrix formulations. 

Only for didactic reasons, we have explained our framework 
for a relatively simple denoising network~\cite{ZZC+17} and 
restricted our experimental comparisons to the classical 
bilateral filter~\cite{AW95,SB97,TM98}. 
However, we also saw that derivative-based concepts, such as 
the popular saliency maps for classification networks~\cite{SVZ14}
or adversarial perturbations~\cite{PMJ+16,SZSB14}, are specific 
instances within our setting.  Putting saliency maps on a solid 
signal processing foundation creates a novel bridge from 
classical model-based signal and image processing to modern 
explainable AI.

Our neural echo framework is very general and does not require
any assumptions on network differentiability. It is equally suited 
for image-to-image and classification networks. They can be 
convolutional or fully connected networks, of feedforward or 
recurrent type. Transformer networks can be analyzed as well. 
The extended state transition matrix characterizes fairly arbitrary 
network actions both from a sender (input) and a receiver 
(output) perspective.

The fact that neural networks require the extension of the filter 
framework of Gaa et al.~\cite{GWFC26} with additive terms
creates an interesting byproduct: It also allows the analysis of 
additional {\em model-based} approaches that have not been 
covered in~\cite{GWFC26}.

We believe that establishing such generalizations, connections, and 
reinterpretations is vital to enable a fruitful exchange of ideas 
from different fields. This can benefit all sides. 
In our ongoing work, we are analyzing more sophisticated 
networks, different image processing tasks, and we are 
performing more exhaustive comparisons to other model-based approaches. 



\bibliographystyle{splncs04}
\bibliography{echo-references}

\end{document}